\documentclass[12pt]{article}

\usepackage{scicite}
\usepackage{graphicx,epic,eepic,epsfig,amsmath,latexsym,amssymb,verbatim,subfigure,color}

\newenvironment{sciabstract}{%
\begin{quote} \bf}
{\end{quote}}

\def\squareforqed{\hbox{\rlap{$\sqcap$}$\sqcup$}}
\def\qed{\ifmmode\squareforqed\else{\unskip\nobreak\hfil
\penalty50\hskip1em\null\nobreak\hfil\squareforqed
\parfillskip=0pt\finalhyphendemerits=0\endgraf}\fi}
\def\endenv{\ifmmode\;\else{\unskip\nobreak\hfil
\penalty50\hskip1em\null\nobreak\hfil\;
\parfillskip=0pt\finalhyphendemerits=0\endgraf}\fi}

\mathchardef\ordinarycolon\mathcode`\:
\mathcode`\:=\string"8000
\def\vcentcolon{\mathrel{\mathop\ordinarycolon}}
\begingroup \catcode`\:=\active
  \lowercase{\endgroup
  \let :\vcentcolon
  }

\newcommand{\nc}{\newcommand}
\nc{\rnc}{\renewcommand}
\nc{\beq}{\begin{equation}}
\nc{\eeq}{{\end{equation}}}
\nc{\beqa}{\begin{eqnarray}}
\nc{\eeqa}{\end{eqnarray}}
\nc{\lbar}[1]{\overline{#1}}
\nc{\bra}[1]{\langle#1|}
\nc{\ket}[1]{|#1\rangle}
\nc{\ketbra}[2]{|#1\rangle\!\langle#2|}
\nc{\braket}[2]{\langle#1|#2\rangle}
\nc{\proj}[1]{| #1\rangle\!\langle #1 |}
\nc{\avg}[1]{\langle#1\rangle}
\nc{\Rank}{\operatorname{Rank}}
\nc{\smfrac}[2]{\mbox{$\frac{#1}{#2}$}}
\nc{\Tr}{\operatorname{Tr}}
\nc{\tr}{\operatorname{Tr}}
\nc{\id}{\operatorname{id}}
\nc{\1}{\openone}
\nc{\ox}{\otimes}
\nc{\dg}{\dagger}
\nc{\dn}{\downarrow}
\nc{\cA}{{\cal A}}
\nc{\cB}{{\cal B}}
\nc{\cC}{{\cal C}}
\nc{\cD}{{\cal D}}
\nc{\cE}{{\cal E}}
\nc{\cF}{{\cal F}}
\nc{\cG}{{\cal G}}
\nc{\cH}{{\cal H}}
\nc{\cI}{{\cal I}}
\nc{\cJ}{{\cal J}}
\nc{\cK}{{\cal K}}
\nc{\cL}{{\cal L}}
\nc{\cM}{{\cal M}}
\nc{\cN}{{\cal N}}
\nc{\cO}{{\cal O}}
\nc{\cP}{{\cal P}}
\nc{\cQ}{{\cal Q}}
\nc{\cR}{{\cal R}}
\nc{\cS}{{\cal S}}
\nc{\cT}{{\cal T}}
\nc{\cX}{{\cal X}}
\nc{\cY}{{\cal Y}}
\nc{\cZ}{{\cal Z}}
\nc{\supp}{{\operatorname{supp}}}
\nc{\var}{\operatorname{var}}
\nc{\rar}{\rightarrow}
\nc{\lrar}{\longrightarrow}
\nc{\polylog}{\operatorname{polylog}}

\nc{\RR}{{{\mathbb R}}}
\nc{\CC}{{{\mathbb C}}}
\nc{\FF}{{{\mathbb F}}}
\nc{\NN}{{{\mathbb N}}}
\nc{\ZZ}{{{\mathbb Z}}}
\nc{\PP}{{{\mathbb P}}}
\nc{\QQ}{{{\mathbb Q}}}
\nc{\UU}{{{\mathbb U}}}
\nc{\EE}{{{\mathbb E}}}
\nc{\Icoh}{{I^c}}
\nc{\Qca}{{Q_{\rm ss}}}
\nc{\Qcaa}{{Q^{(1)}_{\rm ss}}}
\nc{\Dcaa}{{D^{(1)}_{{\rm ss}\rightarrow}}}
\nc{\Dca}{{D_{{\rm ss}\rightarrow}}}

\nc{\be}{\begin{equation}}
\nc{\ee}{{\end{equation}}}
\nc{\bea}{\begin{eqnarray}}
\nc{\eea}{\end{eqnarray}}
\nc{\<}{\langle}
\rnc{\>}{\rangle}
\nc{\Hom}[2]{\mbox{Hom}(\CC^{#1},\CC^{#2})}
\nc{\rU}{\mbox{U}}

\newcounter{lastnote}
\newenvironment{scilastnote}{%
\setcounter{lastnote}{\value{enumiv}}%
\addtocounter{lastnote}{+1}%
\begin{list}%
{\arabic{lastnote}.}
{\setlength{\leftmargin}{.22in}}
{\setlength{\labelsep}{.5em}}}
{\end{list}}

\renewcommand\citeform[1]{#1\,}
\title{Quantum Communication With Zero-Capacity Channels}

\author
{Graeme Smith$^{1}$ and Jon Yard$^{2}$\\
\\
\normalsize{$^{1}$IBM TJ Watson Research Center} \\
\normalsize{1101 Kitchawan Road, Yorktown Heights, NY 10598}\\ \\
\normalsize{$^{2}$Quantum Institute} \\
\normalsize{Center for Nonlinear Studies (CNLS)} \\
\normalsize{Computer, Computational and Statistical Sciences ({CCS-3})}\\
\normalsize{Los Alamos National Laboratory} \\ 
\normalsize{Los Alamos, NM 87545}\\
\\
}

\date{}

\begin{document} 


\baselineskip24pt


\maketitle


\begin{sciabstract}
Communication over a noisy quantum channel introduces errors in the transmission that must be corrected.  A fundamental bound on quantum error correction is the quantum capacity, which quantifies the amount of quantum data that can be protected.  We show theoretically that two quantum channels, each with a transmission capacity of zero, can have a nonzero capacity when used together.  This unveils a rich structure in the theory of quantum communications, implying that the quantum capacity does not uniquely specify a channel's ability for transmitting quantum information.
\end{sciabstract}

% In setting up this template for *Science* papers, we've used both
% the \section* command and the \paragraph* command for topical
% divisions.  Which you use will of course depend on the type of paper
% you're writing.  Review Articles tend to have displayed headings, for
% which \section* is more appropriate; Research Articles, when they have
% formal topical divisions at all, tend to signal them with bold text
% that runs into the paragraph, for which \paragraph* is the right
% choice.  Either way, use the asterisk (*) modifier, as shown, to

\clearpage

Noise is the enemy of all modern communication links.  Cellular, internet and satellite communications all depend crucially on active steps taken 
to mitigate and correct for noise. 
The study of communication in the presence of noise was formalized by Shannon\cite{Shannon48}, 
who simplified the analysis by making probabilistic assumptions about the nature of the noise. 
By modeling a noisy channel $\cN$ as a probabilistic map from input signals to output signals, 
the capacity $\cC(\cN)$ of $\cN$ is defined as the number of bits which can be 
transmitted per channel use, with vanishing errors in the limit of many transmissions.   
This capacity is computed via the formula $\cC(\cN) = \max_X I(X;Y)$ where the maximization is over random variables $X$ at the input of the channel, 
$Y$ is the resulting output of the channel and the mutual information $I(X;Y) = H(X) + H(Y) - H(X,Y)$ 
quantifies the correlation between input and output.  $H(X) = -\sum_x p_x  \log_2 p_x$ denotes the `Shannon entropy', which quantifies the amount of 
randomness in $X$. The capacity, measured in bits per channel use, is the fundamental bound between 
communication rates that are achievable in principle, and those which are not.  
The capacity formula guides the design of practical error 
correction techniques by providing a benchmark against which engineers can test the performance 
of their systems.  Practical implementations guided by the capacity 
result now come strikingly close to the Shannon limit \cite{RU03}.

A fundamental prediction of the capacity formula is that the only channels with zero capacity are precisely those for which the input and 
output are completely uncorrelated.  Furthermore, suppose one is given simultaneous access to two noisy 
channels $\cN_1$ and $\cN_2$.  The capacity of the product channel $\cN_1\times \cN_2$, where the 
channels are used in parallel,  takes the simple 
form $\cC(\cN_1\times \cN_2) = \cC(\cN_1) + \cC(\cN_2)$, i.e., the capacity is additive.  
Additivity shows that capacity is an intrinsic measure of the information conveying properties of 
a channel.

Quantum data is an especially delicate form of information and is particularly 
susceptible to the deleterious effects of noise.  Because quantum communication promises to allow 
unconditionally secure communication\cite{BB84}, and a quantum computer could dramatically 
speed up some computations\cite{Shor94}, there is tremendous interest in 
techniques to protect quantum data from noise.  A quantum channel $\cN$ models a  
physical process which adds noise to a quantum system via an interaction with an unobservable environment (Fig.~1), generalizing Shannon's model and enabling a more accurate depiction of the
 underlying physics.  In this setting, it is natural to ask for the capacity of a 
quantum channel for transmitting quantum mechanical information\cite{BS04}, and 
whether it has a simple formula in analogy with Shannon's.

Just as any classical message can be reversibly expressed as a sequence of bits, a quantum message, i.e.\ an arbitrary state of a given quantum system, can be reversibly transferred to a collection of two-level quantum systems, or `qubits', giving a measure of the size of the system.  The goal of quantum communication is to transfer the joint state of a collection of qubits from one location to another (Fig.~2).
The quantum capacity $\cQ(\cN)$ of a quantum channel $\cN$ is the number of qubits per channel use that can be 
reliably transmitted via many noisy transmissions, where each transmission is modeled by $\cN$.  
Although noiseless quantum communication with a noisy quantum channel is 
one of the simplest and most natural communication tasks one can imagine 
for quantum information, it is not nearly as well understood as 
its classical counterpart.

An analogue for mutual information in the quantum capacity has been proposed \cite{SN96} and called the
`coherent information':  
\begin{equation}
\cQ^{(1)}(\cN) = \max_{\rho^A}\left( H(B) - H(E)\right).
\label{Eq:Ic}
\end{equation}
The entropies are measured on the states induced at the output and environment of the channel (Fig.~2) by the input state $\rho^A$, where $H(B)$ is the `von Neumann entropy' of the state $\rho^B$ at the output. 
Coherent information is rather different from mutual information.  This difference
is closely related to the no-cloning theorem\cite{WZ82}, which states that quantum information cannot be copied,
as the coherent information roughly measures how much more information $B$ holds than $E$.  The no-cloning theorem itself
is deeply tied to the fundamentally quantum concept of entanglement, in which the whole of a quantum system can be in a
definite state while the states of its parts are uncertain.

The best known expression for the quantum capacity $\cQ$ is given\cite{Lloyd97,Shor02,D03} by
the `regularization' of $\cQ^{(1)}$: 
\[\cQ(\cN) = \lim_{n\rightarrow \infty}\frac{1}{n}\cQ^{(1)}(\cN^{\times n}).\] 
Here $\cN^{\times n}$ represents the parallel use of $n$ copies of $\cN$.
The asymptotic nature of this expression prevents one from determining the quantum capacity of a given channel in any effective way, while also making it difficult to reason about its general properties.  
In contrast to Shannon's capacity, where regularization is
unnecessary, here it cannot be removed in general\cite{DSS98,SmithSmo0506}.  
Consequently, even apparently simple questions, such as determining from a 
channel's description whether it can be used to send any quantum information, are currently unresolved.  
We find that the answer to this question depends on context; 
there are pairs of zero-capacity channels which, used together,
 have a positive quantum capacity (Fig.~3).  This shows the quantum capacity is not 
additive, and thus  the quantum capacity of a channel does not completely 
specify its capability for transmitting quantum information.

While a complete characterization of zero-capacity channels is unknown, certain classes of zero-capacity 
channels are known.  One class consists of channels for which the joint quantum state of the 
output and environment is symmetric under interchange.  These `symmetric channels' are quite 
different from Shannon's zero-capacity channels, as they display correlations between the input 
and output.  However, they are useless by themselves for quantum communication because their symmetry implies that 
any capacity would lead to a violation of the no cloning theorem\cite{WZ82,BDS97}.  Another class of 
zero-capacity channels are entanglement-binding channels\cite{HHH96,H97}, also called `Horodecki channels', which can 
only produce very weakly entangled states 
satisfying a condition called positive partial transposition\cite{Peres96}.

Even though channels from one or the other of these classes cannot be combined 
to faithfully transmit quantum data, we find that when one combines a channel 
from each class, it is sometimes possible to obtain a positive quantum capacity.  We do this
by proving a new relationship between two further capacities of a quantum channel:
the private capacity\cite{D03} and the assisted capacity\cite{SSW06}.

The private capacity $\cP(\cN)$ of a quantum channel $\cN$ is the rate at which
it can be used to send classical data that is secure against
an eavesdropper with access to the environment of the channel.
 This capacity is closely related to quantum key distribution protocols\cite{BB84}
and was shown\cite{D03} to equal the 
 regularization of the `private information':
\begin{equation}\label{Eq:Pee1}
\cP^{(1)}(\cN) = \max_{X, \rho^A_x}\left( I(X;B)-I(X;E)\right),
\end{equation}
where the maximization is over classical random variables $X$ and quantum states $\rho^A_x$ on the input of $\cN$ depending on the value $x$ of $X$.

In order to find upper bounds on the quantum capacity, an `assisted capacity' 
was recently introduced\cite{SSW06} where one allows the free use of arbitrary
 symmetric channels to assist quantum communication over a given channel. 
Letting $\cA$ denote a symmetric channel of unbounded dimension (the strongest such channel), 
the assisted capacity $\cQ_\cA(\cN)$ of a quantum channel $\cN$ satisfies\cite{SSW06} 
\[\cQ_{\cA}(\cN) = \cQ(\cN\times \cA) = \cQ^{(1)}(\cN\times \cA).\]
Because the dimension of the input to $\cA$ is unbounded, we cannot evaluate the assisted capacity in
general.  Nonetheless, the assisted capacity helps to reason about finite-dimensional channels.

While Horodecki channels have zero quantum capacity, examples of such channels
with nonzero private capacity are known\cite{HHHO03,HPHH05}.  
One of the two zero-capacity channels we will combine to give positive joint capacity is such a
`private Horodecki channel' $\cN_H$, and the other is the symmetric channel $\cA$.
Our key tool is the following new relationship between the capacities of any channel 
$\cN$ (Fig.~4):
\begin{equation}\label{Eq:main}
{\mbox{ $\frac{1}{2}$}}\cP(\cN) \leq \cQ_\cA(\cN).
\end{equation}
A channel's assisted capacity is at least as large as half its 
private capacity.
It follows that any private Horodecki channel $\cN_H$ 
has a positive assisted capacity, and thus the two zero-capacity channels $\cN_H$ and $\cA$ satisfy 
\[Q_\cA(\cN_H) = \cQ(\cN_H \times \cA) > 0.\]

Although our construction involves systems of unbounded dimension, one can show that any private Horodecki channel can be combined with a finite symmetric channel to give positive quantum capacity.
In particular, there is a private Horodecki channel acting on a four-level 
system\cite{HPHH05}.  This channel gives positive quantum capacity when combined with a 
small symmetric channel -- a $50\%$ erasure channel $\cA_{e}$ with a four-level input
which half of the time delivers the input state
to the output, otherwise telling the receiver that an erasure has occurred.
We show \cite{NumNote} that the parallel combination of these channels has a quantum capacity greater than  0.01.

We find this `superactivation' to be a startling effect.  One would think that the
question, ``can this communication link transmit any information?" would 
have a straightforward answer.  However, with quantum data, the answer may well be ``it depends on
the context".   Taken separately, private Horodecki channels and symmetric channels are useless
 for transmitting quantum information, albeit for entirely different reasons.  
Nonetheless, each channel has the potential to ``activate" the other, effectively
 canceling the other's reason for having zero capacity.  We know of no analogue of 
this effect in the classical theory.  Perhaps each channel transfers some 
different, but complementary kind of quantum information.  If so, can these 
kinds of information be quantified in an operationally meaningful way?  
Are there other pairs of zero-capacity channels displaying this effect?  
Are there triples?  Does the private capacity also display superactivation?
Can all Horodecki channels be superactivated, or just those with positive private capacity?
What new insights does this yield for computing the quantum capacity in general?  

Besides additivity, our findings resolve two open questions about the quantum capacity.  
First we find \cite{NumNote} that the quantum capacity is not a convex function of the channel.  Convexity of a capacity
means that a probabilistic mixture of two channels never has a higher capacity than the 
corresponding average of the capacities of the individual channels.  Violation of convexity leads to a counterintuitive situation where it can be beneficial to forget which channel is being used.  
We also find \cite{NumNote} channels with an arbitrarily large gap between $\cQ^{(1)}$ -- the so-called `hashing rate'\cite{Lloyd97,Shor02,D03} -- 
and the quantum capacity.  It had been consistent with previous results\cite{DSS98,SmithSmo0506} to believe that $\cQ$ and $\cQ^{(1)}$ would be equal up to small corrections. Our work shows this is not the case and indicates that the hashing rate is an overly pessimistic benchmark against which to measure the performance of practical error-correction schemes.  This could be good news for the analysis of fault tolerant quantum computation in the very noisy regime.

Forms of this sort of superactivation are known in the multiparty setting, where
several separated parties communicate via a quantum channel with multiple 
inputs or outputs\cite{SST03,DCH04,DS07,CH08}, and have been conjectured for
a quantum channel assisted by
classical communication between the sender and receiver\cite{HHH99,SST01}.  
Because these settings are rather complex, it is perhaps 
unsurprising to find exotic behavior.  In contrast, the problem of noiseless quantum communication with
a noisy quantum channel is one of the simplest and most natural communication tasks imaginable in a 
quantum mechanical context.  Our findings uncover a level of complexity in this simple problem that had not
been anticipated and point towards several fundamentally new questions about information and communication
in the physical world.

% Following is a new environment, {scilastnote}, that's defined in the
% preamble and that allows authors to add a reference at the end of the
% list that's not signaled in the text; such references are used in
% *Science* for acknowledgments of funding, help, etc.

% For your review copy (i.e., the file you initially send in for
% evaluation), you can use the {figure} environment and the
% \includegraphics command to stream your figures into the text, placing
% all figures at the end.  For the final, revised manuscript for
% acceptance and production, however, PostScript or other graphics
% should not be streamed into your compliled file.  Instead, set
% captions as simple paragraphs (with a \noindent tag), setting them
% off from the rest of the text with a \clearpage as shown  below, and
% submit figures as separate files according to the Art Department's
% instructions.

\clearpage

%\bibliography{Super}

\bibliographystyle{Science}

\begin{scilastnote}
\item We are indebted to Charlie Bennett, Clifton Callaway, Eddy Timmermans, 
Ben Toner and Andreas Winter for encouragement and comments on an earlier draft.  
JY is supported by the Center for Nonlinear Studies (CNLS) and the Quantum Institute 
through grants provided by the LDRD program of the U.S. Department of Energy.
\end{scilastnote}

\clearpage
\begin{figure}
\includegraphics[width=16cm]{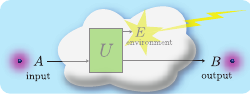}
\caption{Representation of a quantum channel.  A channel reversibly transfers the state of a physical system in the laboratory of the sender to the combination of the system possessed by the receiver and an `environment' which is inaccessible to the users of the channel.  Discarding the environment results in a noisy evolution of the state.  The input and output denote separate places in space and or time, modeling for example a leaky optical fiber or the irreversible evolution of the state of a quantum dot.}
\end{figure}

\begin{figure}
\includegraphics[width=16cm]{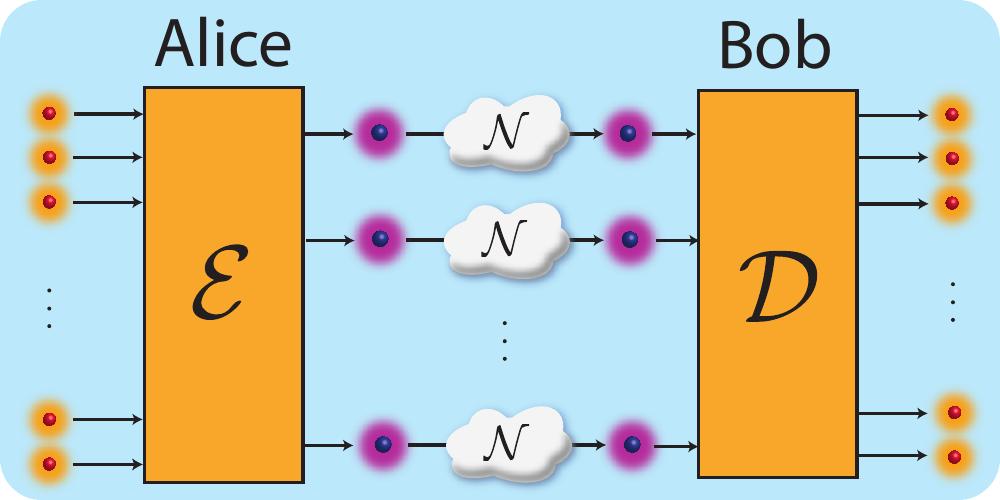}
\caption{The quantum capacity of a quantum channel.  Quantum data is held by a sender (traditionally 
called Alice), who would like to transmit it to a reciever (Bob) with many parallel uses of a noisy
quantum channel $\cN$.  Alice encodes the data with a collective encoding operation $\cE$ which results
in a joint quantum state on the inputs of the channels $\cN^{\times n}$.  The encoded state is sent through
the noisy channels.  When Bob receives the state, he applies a decoding operation $\cD$ which acts 
collectively on the many outputs of the channels.  After decoding, Bob holds the state which Alice wished
to send.  The quantum capacity is the total number of qubits in the state Alice sends divided by the number
of channel uses.}
\end{figure}

\begin{figure}
\includegraphics[width=16cm]{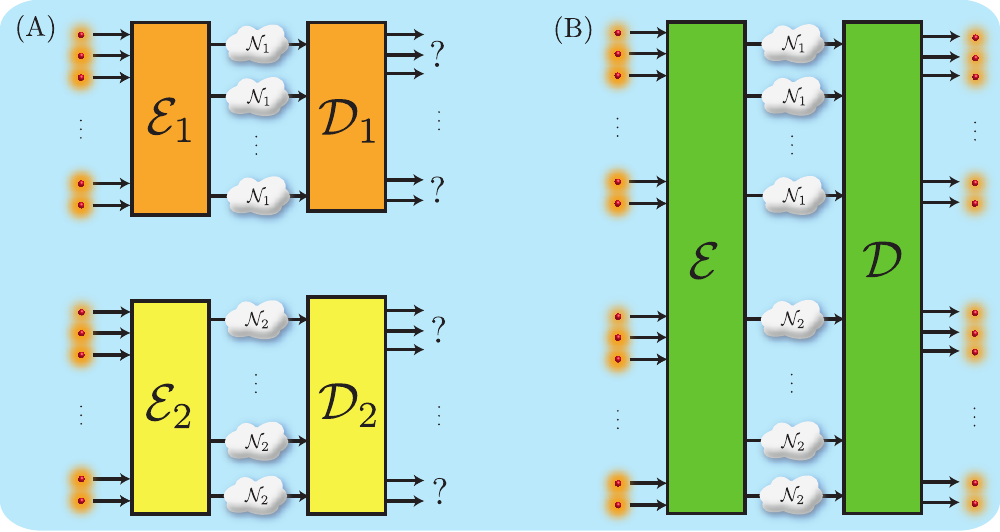}

\caption{(A) Alice and Bob attempt to separately use two zero-capacity channels $\cN_1$ and $\cN_2$ to transfer quantum states.  Alice uses separate encoders $\cE_1$ and $\cE_2$ for each group of channels and Bob uses separate decoders $\cD_1$ and $\cD_2$.  Any attempt will fail because the capacity of each channel is zero.  (B) The same two channels being used in parallel for the same task.  Alice's encoder $\cE$ now has simultaneous access to the inputs of all channels being used and Bob's decoding $\cD$ is also performed jointly.  Noiseless communication is nonetheless possible because $\cQ$ is not additive. }
\end{figure}

\begin{figure}
\includegraphics[width=16cm]{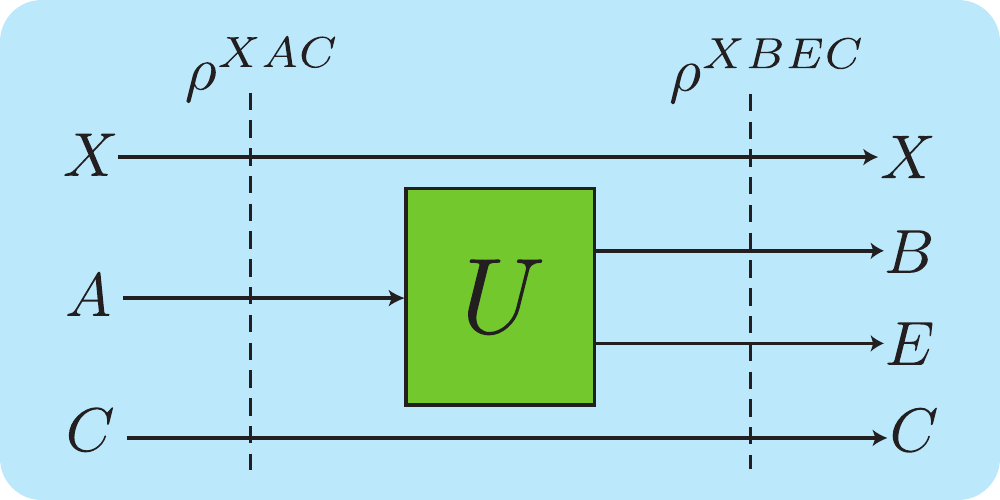}
\caption{Relating the private capacity and the assisted capacity.  
A straightforward proof of Eq.~\ref{Eq:main} uses the expression\cite{SSW06}  
 $\cQ_\cA(\cN) = \frac{1}{2}\max_{\rho^{XAC}}( I(X;B|C)- I(X;E|C)).$
Here, $I(X;B|C)$ is the `conditional mutual information': $H(XC)+H(BC)-H(XBC)-H(C)$. 
It is evaluated on the state obtained by putting the $A$ part of a state $\rho^{XAC}$ into the channel $\cN$, which can be thought of as mapping $A\to BE$ as in Fig.~2. 
The maximization here is similar in form to Eq.~\ref{Eq:Pee1}, but 
is over a less constrained type of state.  Therefore,   
$\frac 12 \cP^{(1)}(\cN) \leq \cQ_\cA(\cN)$.  This bound holds for the associated regularized 
quantities and since regularization does not change $\cQ_\cA$, Eq.~\ref{Eq:main} follows.}
\end{figure}

\clearpage

\clearpage

%%%%%%%%%%%%%%%%%%%%%%%%%%%%%%%
% SOM
%%%%%%%%%%%%%%%%%%%%%%%%%%%%%%%

\setcounter{page}{1}
\setcounter{equation}{0}
\renewcommand{\theequation}{S\arabic{equation}}

\begin{center}
\Large{Quantum Communication with Zero-Capacity Channels} \\ 
\vspace{.2in}
\normalsize{(Supporting Online Material)} \\
\vspace{.2in}
\large{Graeme Smith and Jon Yard}
\end{center}

In this supporting material, we assume a basic knowledge of quantum information theory at the level of \cite{NC00S}.  We denote the dimension of a Hilbert space $A$ as $|A|$ while abbreviating tensor products as $AB = A\ox B$ and the restriction of a density matrix $\rho^{AB}$ to the subsystem $A$ as $\rho^A$.  By an ensemble of states $\{p_x,\rho^A_x\}$, we mean that the density matrix $\rho^A_x$ on $A$ occurs with probability $p_x$.  We freely associate such ensembles with the joint state $\sum_x p_x \proj{x}^X\ox \rho_x^A$, where $\{\ket{x}^X\}$ denotes an orthonormal basis for $X$.   For a channel with input $A$, output $B$ and environment $E$, we abbreviate $I_{c}(\cN,\rho^A) = H(B) - H(E)$ so that 
$\cQ^{(1)}(\cN) = \max_{\rho^A} I_c(\cN,\rho^A)$.

\begin{center}
{\bf Superactivation with finite dimensional channels}\\
\end{center}
Our key tool in this work was the relationship $\frac 12 \cP(\cN) \leq \cQ(\cN \otimes \cA) = \cQ_{\cA}(\cN)$, which is valid for any quantum channel $\cN$.  A disadvantage of this result is that the input and output systems of the channel $\cA$ are infinite.   Guided by that result, we now give a weaker, more manageable bound in which a $50\%$-erasure channel $\cA_e$ with finite-dimensional input and output systems plays the role of $\cA$.

Consider a channel $\cN$ with input $A$, output $B$ and environment $E$.  By Eq.~\ref{Eq:Pee1} in the main text, observe that every ensemble $\{p_x,\rho^A_x\}$ yields the lower bound 
\[I(X;B) - I(X;E) \leq \cP^{(1)}(\cN)\]
where the mutual informations are evaluated on the state 
\[\rho^{XBE} = \sum_x p_x \proj{x}^X\ox \rho_x^{BE}\]
and $\rho_x^{BE}$ is the joint state of the output and environment when $\rho^A_x$ is sent through the channel.  
Furthermore, if the input to $\cN$ is finite, this bound is achievable using a finite ensemble \cite{D03S}.

Currently, every known private Horodecki channel $\cN_H$ also satisfies $\cP^{(1)}(\cN_H) > 0$.  Therefore, for such a channel, there is a finite ensemble for which $I(X;B)-I(X;E)>0$.  With this in mind, we will demonstrate the following: \\ \\
{\bf Theorem:} 
{\it Given an ensemble $\{p_x,\rho_x^A\}$ and a channel $\cN$ with input $A$, output $B$ and environment $E$, let $\cA_e$ be a  50\%-erasure channel with input space $C$ of dimension equal to the sum of the ranks of the states $\rho_x^A$.  Then there is a state $\rho^{AC}$ such that } 
\[I_c(\cN\ox \cA_e,\rho^{AC}) = \smfrac 12 \big(I(X,B) - I(X;E)\big).\]
%\[\cP^{(1)}(\cN) \leq \cQ^{(1)}(\cN\ox \cA_e).\]
The next section describes a private Horodecki channel\cite{HPHH05S} $\cN^{(4)}_H$ with a four-dimensional input and an ensemble with two rank-two states such that $I(X;B) - I(X;E) > 0.02$.  Therefore, there is an erasure channel $\cA_e^{(4)}$ with a four-dimensional input $C$ and a state $\rho^{AC}$ such that $\cQ(\cN_H\ox \cA_e) \geq  I_c(\cN\ox \cA_e,\rho^{AC})> 0.01$. 

Our strategy of proof is as  follows. When the states in the ensemble are pure, i.e.\ $\rho_x^A = \proj{\rho_x}^A$, we have the identity \cite{D03S} 
\[I_c(\cN,\rho^A) = I(X;B) - I(X;E)\] where $\rho^A = \sum_x p_x \proj{\rho_x}$.  Consequently, the coherent information is obtained by restricting the maximization for the private information to pure state ensembles.  Since we begin with a mixed state ensemble, we consider a related ensemble of purified states and send the purifying system through a 50\%-erasure channel. \\ \\
{\bf Proof of Theorem:}  Define purifications $\ket{\rho_x}^{AC}$ of the states $\rho_x^A$ such that the supports of the $\rho_x^C$ are disjoint.
Then the pure state
\[\ket{\rho}^{XAC} = \sum_x \sqrt{p_x} \ket{x}^{X}\ket{\rho_x}^{AC}\]
is a purification of the state $\sum_x p_x \proj{x}^X\ox \rho^A_x$ associated to the ensemble.
We will evaluate the coherent information resulting from sending $A$ through $\cN$ and $C$ through $\cA_e$.  Denoting the output of $\cA_e$ by $D$ and the environment of $\cA_e$ by $F$, we obtain the following chain of inequalities:
\begin{eqnarray}
I_c(\cN\otimes \cA_e,\rho^{AC}) &=& H(BD) - H(EF) \label{Eq:chain1} \\
&=& \smfrac 12\big(H(B) - H(EC)\big) + \smfrac 12\big(H(BC) - H(E)\big) \label{Eq:chain2}\\
&=& \smfrac 12\big(H(B) - H(XB)\big) + \smfrac 12\big(H(XE) - H(E)\big) \label{Eq:chain3}\\
&=& \smfrac 12 \big(I(X;B) - I(X;E)\big).\label{Eq:chain4}
\end{eqnarray}
In Eq.~\ref{Eq:chain1}, the entropies are evaluated on the state obtained by sending the $AC$ parts of $\ket{\rho}^{XAC}$ through the respective channels.  Eq.~\ref{Eq:chain2} holds because with equal probability, $\cN_e$ either delivers $C$ to its output $D$ or to its environment $F$, so this difference of entropies can be rewritten in terms of quantities evaluated on the state on $XBEC$ obtained by sending only the $A$ part of $\ket{\rho}^{XAC}$ through $\cN$.   Eq.~\ref{Eq:chain3} is true because bipartitions of any pure state have the same entropy and Eq.~\ref{Eq:chain4} uses the definition of mutual information after adding $H(X)$ to the first term and subtracting it from the second.  \qed

\newpage
\begin{center}
{\bf A four-dimensional private Horodecki channel}
\end{center}
For convenience we give an explicit description of the four-dimensional private Horodecki channel $\cN_H^{(4)}$ from \cite{HPHH05S}.  
The action of any channel $\cN$ can be written in Kraus form as
\begin{equation}
\cN(\rho) = \sum_k N_k \rho N_k^\dg, \nonumber
\end{equation}
where the Kraus matrices $N_k$ satisfy $\sum_k N_k^\dg N_k = I$.
We denote the input of $\cN_H^{(4)}$ as a tensor product of two qubits $A = A_1A_2$ and denote the output as $B$.  This channel is specified by the following six Kraus matrices: 
\[\sqrt{\smfrac q2} I \ox \proj{0},
\sqrt{\smfrac q2}Z \ox \proj{1},
\sqrt{\smfrac q4} Z \ox Y,
\sqrt{\smfrac q4} I \ox X,
\sqrt{1-q} X \ox M_0,
\sqrt{1-q} Y \ox M_1.\]
Here, $q = \frac{\sqrt{2}}{1+\sqrt{2}}$, while $X$,$Y$ and $Z$ are the usual Pauli matrices and 
\begin{eqnarray}\nonumber
M_0 =  \begin{pmatrix}
\frac{1}{2}\sqrt{2+\sqrt{2}} & 0 \\ 0 &\frac{1}{2}\sqrt{2-\sqrt{2}}  
\end{pmatrix},\,\,\,\,\,\,
 M_1  =  \begin{pmatrix}\frac{1}{2}\sqrt{2-\sqrt{2}} & 0 \\ 0 &\frac{1}{2}\sqrt{2+\sqrt{2}} \end{pmatrix}.
\end{eqnarray}
A lower bound of $0.02$ on the private capacity $\cP(\cN_H^{(4)})$ is obtained via the ensemble consisting of two equiprobable states 
$\rho_x^{A} = \proj{x}^{A_1}\ox \frac{1}{2}I^{A_2}$
because the state $\rho^{XBE}$ resulting from putting $A$ into the channel $\cN_H^{(4)}$ satisfies \cite{HPHH05S}
\[I(X;B) - I(X;E) \geq 1 - q \log_2 q - (1-q)\log_2(1-q) > 0.02.\]

\newpage 

\begin{center}
{\bf Nonconvexity of quantum capacity}
\end{center}
We now use the results of the first section
 to show that $\cQ$ is not convex.  Fix a private Horodecki channel $\cN_H$ and a 50\%-erasure channel $\cA_e$ such that the input $A$ to $\cN_H$ and the input $C$ to $\cA_e$ have the same dimension, and such that there is a state $\rho^{AC}$ which is symmetric under interchanging $A$ and $C$ satisfying $I_c(\cN_H\ox \cA_e,\rho^{AC}) > 0$.  In particular, the channel $\cN_H^{(4)}$ from the previous section and a four-dimensional erasure channel $\cA_e^{(4)}$ satisfy these criteria with respect to the state 
 \[\rho^{AC} = \frac 12 \big(\proj{0}^{A_1}\ox \proj{0}^{C_1}  +\proj{1}^{A_1}\ox \proj{1}^{C_1}\big) \ox \proj{\phi_+}^{A_2C_2}\]
where $\ket{\phi_+} = \frac{1}{\sqrt{2}}(\ket{00} + \ket{11}),$ $A = A_1A_2$ and $C = C_1C_2$.
Identifying $A\simeq C$,  we define the channel 
\[\cM_p = p\cN_H \ox \proj{0} + (1-p) \cA_e \ox \proj{1}\]
where $0\leq p \leq 1$.  
With probability $p$, this channel applies $\cN_H$ to the input and otherwise applies $\cA_e$, while telling the receiver which channel was applied.   
Although $\cM_p$ is a convex combination of the zero-capacity channels $\cN_H \ox \proj{0}$ and $\cA_e\ox \proj{1}$, we will show that for small enough values of $p$, their convex combination $\cM_p$ has a positive capacity.  On any input state $\rho$,  we have 
\begin{eqnarray*}
I_c(\cM_p\ox \cM_p,\rho) &=& 
p^2I_c(\cN_H\ox \cN_H,\rho)
 + p(1-p)I_c(\cN_H\ox \cA_e,\rho)  \\
& & +\, p(1-p)I_c(\cA_e\ox \cN_H,\rho)
+ (1-p)^2I_c(\cA_e\ox \cA_e,\rho). 
\end{eqnarray*}
Since $\cA_e \ox \cA_e$ is a symmetric channel, the last term is always zero.  
Choosing the state $\rho = \rho^{AC}$, which was assumed to be symmetric,
we find that
\begin{equation}
I_c(\cM_p \ox \cM_p,\rho^{AC}) = 2p(1-p)I_c(\cN_H \ox \cA_e,\rho^{AC}) + p^2I_c(\cN_H\ox \cN_H,\rho^{AC}).\nonumber
\end{equation}
For $0 < p < 1$, the first term is positive by assumption.  The second term can never be greater than zero because $\cQ(\cN_H) = 0$, although it is lower bounded by $-2p^2c$, where $c = \log_2 |E|$.  Here $c$ is finite because $|E|$ is finite when the input and output of $\cN_H$ have finite dimension.     
Simple algebra reveals that $I_c(\cM_p \ox \cM_p,\rho^{AC}) >0$ for all $p$ satisfying 
\[0 < p < \frac{I_c(\cN_H \ox \cA_e,\rho^{AC})}{c + I_c(\cN_H \ox \cA_e,\rho^{AC})}.\]
For the four-dimensional example of the previous section, one has $c = \log_2 6$ so the corresponding convex combination has a positive capacity if $0 < p < 0.0041$.  Stronger violations are expected to be found in larger examples.

\begin{center}
{\bf Arbitrarily large gap between $\cQ^{(1)}$ and $\cQ$}
\end{center}
Although it has long been known that $\cQ$ can be strictly greater than $\cQ^{(1)}$, there has been speculation that deviations of $\cQ$ from $\cQ^{(1)}$
may be fairly small.  Thus, while the regularized nature of the capacity expression is unwieldy, 
we might hope that for practical purposes the quantum capacity is well approximated by $\cQ^{(1)}$ and
analysis could proceed by considering the computable function $\cQ^{(1)}$.  Our work shows that this is not true, as there exist channels $\cM$ with $\cQ^{(1)}(\cM) = 0$ for which the actual capacity
can be arbitrarily large.   Let $\cN_H$ be a private Horodecki channel and let $\cA_e$ be a 50\%-erasure channel with the same input dimension 
for which  $\cQ^{(1)}(\cN_H\ox \cA_e) > 0$. For example, the four-dimensional channels discussed above would work.  Define $\cM$ to be a channel with input $A = A_1A_2$, where $A_1$ is a qubit and $A_2$ is the input space of $\cN_H$ and $\cA_e$. 
The channel measures the first qubit $A_1$ in the $\{\ket{0},\ket{1}\}$ basis and, depending on the outcome, applies one of the channels $\cN_H$ or $\cA_e$ to $A_2$.   The outcome of the measurement is revealed to the receiver.  
This channel can be seen to have $\cQ^{(1)}(\cM) = 0$ because
$\cQ^{(1)}(\cN_H) = \cQ^{(1)}(\cA_e) = 0$.  
However, the sender has control over which channel is applied to which input, so
$\cQ^{(1)}(\cM\ox\cM) \geq \cQ^{(1)}(\cN_H \ox \cA_e)>0$.  By replacing $\cN_H$ in the above discussion with $n$ instances
of $\cN_H$, and similarly for $\cA_e$, this violation can be made arbitrarily large.

%%%%%%%%%%%%%%%%%%%%%%%%%%%%%%%%%%
% END OF SOM
%%%%%%%%%%%%%%%%%%%%%%%%%%%%%%%%%%

%\bibliographystyle{Science}

\end{document}